\documentclass[prd,showpacs,showkeys,nofootinbib,twocolumn]{revtex4-1}

\usepackage{graphicx}
\usepackage{amsmath,amsfonts,amssymb,amsxtra,latexsym}

\usepackage{hyperref}
\hypersetup{
    colorlinks=true,
    linkcolor=red,
    citecolor=blue,      
    urlcolor=blue,
}

\usepackage[utf8]{inputenc}

\begin{document}

\title{Gravitational Waves with Colliding or Non--Colliding Wave Fronts}

\author{Peter A. Hogan}
\email{peter.hogan@ucd.ie}
\affiliation{School of Physics, University College Dublin, Belfield, Dublin 4, Ireland} 

\author{Dirk Puetzfeld}
\email{dirk.puetzfeld@zarm.uni-bremen.de}
\homepage{http://puetzfeld.org}
\affiliation{University of Bremen, Center of Applied Space Technology and Microgravity (ZARM), 28359 Bremen, Germany} 

\date{ \today}

\begin{abstract}
The known exact solutions of Einstein's vacuum field equations modeling the gravitational fields of pure gravitational radiation involve wave fronts which are either planar or roughly spherical. We describe a scheme designed to check explicitly whether or not the wave fronts collide. From the spacetime point of view the scheme determines whether or not the null hypersurface histories of the wave fronts intersect and, in particular, allows easy identification of the cases in which the null hypersurfaces do not intersect.
\end{abstract}

\pacs{04.20.-q; 04.20.Jb; 04.20.Cv}
\keywords{Classical general relativity; Exact solutions; Fundamental problems and general formalism}

\maketitle


\section{Introduction} \label{sec:introduction}

Arguably the simplest exact models in general relativity of the gravitational field due to gravitational radiation are the exact solutions of Einstein's vacuum field equations describing the gravitational field due to a system or train of gravitational waves having clearly identifiable wave fronts which are either plane (non--expanding) or spherical (in the sense that the wave fronts are compact and expanding). The histories of these wave fronts in spacetime are shear--free null hypersurfaces. Important from a physical point of view is whether or not the wave fronts in these models do or do not collide. From the space--time point of view the issue is whether or not the null hypersurface histories of the wave fronts \emph{intersect}. 

The object of this paper is to provide a scheme to determine this property of the wave fronts initially for the case of Kundt waves compared to the more special plane fronted waves with parallel rays (the so--called pp--waves)  \cite{Kundt:1961, Ehlers:Kundt:1962, Pirani:1964:1, Stephani:etal:2003}. This property of plane fronted waves has been known for a long time. For example Pirani \cite{Pirani:1964:1} described it impressively when he wrote: ``If one thinks of a plane wave as represented by the parallel beam from a searchlight which points in a certain direction, then a plane fronted wave may arise if the searchlight turns about and its beam sweeps back and forth across the sky". This electromagnetic analogy is described explicitly in Appendix \ref{app_A} (making use of the geometrical construction given in section \ref{sec:planewaves} below). The histories of plane fronted waves in spacetime are null hypersurfaces which are shear--free and expansion--free (i.e.\ null hyperplanes). We give here a geometrical construction of Kundt waves which illustrates explicitly how the null hypersurface histories of the waves \emph{intersect}, in contradistinction to the pp--waves for which the histories in spacetime \emph{do not intersect}. A similar geometrical construction has been utilized in \cite{Hogan:2018} for the construction of plane fronted waves in the presence of a cosmological constant \cite{Ozsvath:etal:1985}. The gravity waves with expanding wave fronts propagating in a vacuum are the Robinson--Trautman \cite{Robinson:Trautman:1960,Robinson:Trautman:1962} purely radiative solutions of Einstein's vacuum field equations. The histories of such waves in spacetime are null hypersurfaces which are shear--free and expanding. We extend the geometrical construction of the plane fronted case to the case of expanding null hypersurfaces \emph{which may or may not intersect}. Our starting point in both the non--expanding and expanding wave cases will involve shear--free null hypersurfaces in Minkowskian spacetime, and we utilize the fact that the only shear--free null hypersurfaces in Minkowskian spacetime are null hyperplanes or null cones or portions thereof \cite{Penrose:1972}. 

\section{Plane Fronted Waves} \label{sec:planewaves}

The line element of Minkowskian spacetime in rectangular Cartesian coordinates and time $X^i=(X, Y, Z, T)$ with $i=1, 2, 3, 4$, is
\begin{equation}\label{1}
ds_0^2=(dX)^2+(dY)^2+(dZ)^2-(dT)^2=\eta_{ij}\,dX^i\,dX^j\ .
\end{equation}
The null hyperplanes in these coordinates are given by $u(X, Y, Z, T)={\rm constant}$, with $u(X, Y, Z, T)$ given implicitly by the equation
\begin{equation}\label{2}
\eta_{ij}\,a^i(u)\,X^j+n(u)=0\ \ \ {\rm with}\ \ \ \eta_{ij}\,a^i\,a^j=0\ ,
\end{equation}
and $n(u)$ an arbitrary real--valued function of $u$ subject to $\gamma(u)=\dot n(u)\neq 0$. Here and throughout a dot will indicate differentiation with respect to $u$. Since $\gamma\neq 0$ only the \emph{direction} of $a^i$ is involved in (\ref{2}) since any function of $u$ multiplying $a^i$ can be absorbed into $n(u)$. We thus parametrize $a^i(u)$ with a complex--valued function $l(u)$ and its complex conjugate $\bar l(u)$ as
\begin{equation}\label{3}
a^1+i\,a^2=2\sqrt{2}\,l\ , a^3+a^4=4\,l\,\bar l\ , a^3-a^4=-2\ .
\end{equation}
The partial derivative of (\ref{2}) with respect to $X^j$ (indicated by a comma) results in 
\begin{equation}\label{4}
u_{,j}=-\frac{a_j}{\gamma+\dot a_i\,X^i}\ ,
\end{equation}
with $a_i=\eta_{ij}\,a^j$. This confirms that $u={\rm constant}$ are null hypersurfaces. For confirmation that $u={\rm constant}$ are shear--free and expansion--free see \cite{Hogan:2018} (they are obviously twist--free). With the parametrization (\ref{3}) the equation (\ref{2}) reads 
\begin{equation}\label{5}
Z+T=\sqrt{2}\,\bar l\,(X+i\,Y)+\sqrt{2}\,l\,(X-i\,Y)+2\,l\,\bar l\,(Z-T)+n\ .
\end{equation}
It is convenient to introduce a complex coordinate $\zeta$ given by
\begin{equation}\label{6}
\zeta=\frac{1}{\sqrt{2}}\,(X+i\,Y)+l\,(Z-T)\ ,
\end{equation}
which results in 
\begin{eqnarray}
X+i\,Y&=&\sqrt{2}\,\zeta-\sqrt{2}\,l\,(Z-T)\ ,\label{7}\\
Z+T&=&2\,(\bar l\,\zeta+l\,\bar\zeta)-2\,l\,\bar l\,(Z-T)+n\ .\label{8}
\end{eqnarray}
Making these substitutions in the Minkowskian line element (\ref{1}) results in 
\begin{equation}\label{9}
ds_0^2=2\,|d\zeta-\beta(u)\,(Z-T)\,du|^2+2\,q\,du\,(dZ-dT)\ ,
\end{equation}
with $\beta(u)=\dot l(u)$ and 
\begin{equation}\label{10}
q(\zeta, \bar\zeta, u)=\bar\beta\,\zeta+\beta\,\bar\zeta+\frac{1}{2}\,\gamma\ .
\end{equation}
Using (\ref{7}) and (\ref{8}) in (\ref{4}) we arrive at
\begin{equation}\label{11}
a_j=-2\,q\,u_{,j}\ \ \Leftrightarrow\ \ a_j\,dX^j=-2\,q\,du\ .
\end{equation}
If we form the Kerr--Schild \cite{Kerr:Schild:1965} metric $g_{ij}=\eta_{ij}+W\,(X, Y, Z, T)\,a_i\,a_j$, for some real--valued function $W$, then $a^i=\eta^{ij}\,a_j=g^{ij}\,a_j$ is null, geodesic, shear--free, expansion--free and twist--free in the spacetime with metric $g_{ij}$. Also $g_{ij}\,a^i\,X^j+n=\eta_{ij}\,a^i\,X^j+n=0$ by (\ref{2}) and so the equation for $u(X, Y, Z, T)$ is identical in Minkowskian spacetime and in the spacetime with metric $g_{ij}$. Writing the Kerr--Schild line element in coordinates $\zeta, \bar\zeta, Z-T=v$ and $u$, using (\ref{9}), we have
\begin{equation}\label{12}
ds^2=2\,|d\zeta-\beta(u)\,v\,du|^2+2\,q\,du\,(dv+F\,du)\ ,
\end{equation}
with $F=F(\zeta, \bar\zeta, v, u)$ and we have written $2\,q\,W=F$ for convenience. This form of line element involving the functions $\beta$ and $q$ is a unique consequence of our geometrical construction which is the origin of these functions. A direct comparison between the line element (\ref{12}) and the Kundt line element is given in section \ref{sec:discussion} below. Now Einstein's vacuum field equations require $F$ to take the form $F=A(u)\,v+B(\zeta, \bar\zeta, u)$ and, in addition, $F$ must satisfy the field equation
\begin{equation}\label{13}
\frac{\partial^2F}{\partial\zeta\partial\bar\zeta}=0\ .
\end{equation}
Replacement of the coordinate $v$ by $v'=v+S(\zeta, \bar\zeta, u)$ with
\begin{equation}\label{13'}
S=-\frac{1}{2}q\,(\beta\,\bar\beta)^{-1}A(u)\ \ \ \Rightarrow\ \ \frac{\partial S}{\partial\zeta}=q^{-1}\bar\beta\,S\ ,
\end{equation}
preserves the form of the line element (\ref{12}) while eliminating the term involving the coordinate $v$ in $F$ so that, in effect, we can take $F=F(\zeta, \bar\zeta, u)$. The Newman--Penrose \cite{Newman:Penrose:1962} components $\Psi_A$ with $A=0, 1, 2, 3, 4$, of the Riemann curvature tensor vanish with the exception of 
\begin{equation}\label{14}
\Psi_4=q^{-1}\frac{\partial^2F}{\partial\zeta\partial\zeta}\ .
\end{equation}
This curvature tensor is Petrov Type N (i.e.\ purely radiative) with degenerate principal null direction, in coordinates $x^{i'}=(\zeta, \bar\zeta, v, u)$, given by the vector field or differential 1--form
\begin{equation}\label{15}
k^{i'}\,\frac{\partial}{\partial x^{i'}}=\frac{\partial}{\partial v}\ \ {\rm and}\ \ k_{i'}\,dx^{i'}=q\,du\ .
\end{equation}
In terms of the null tetrad $m^{i'}, \bar m^{i'}, k^{i'}, l^{i'}$  given by (\ref{15}) and
\begin{eqnarray}
m^{i'}\frac{\partial}{\partial x^{i'}}&=&\frac{\partial}{\partial\bar\zeta}\ ,\ \bar m^{i'}\frac{\partial}{\partial x^{i'}}=\frac{\partial}{\partial\zeta}\ \ \ {\rm and}\nonumber\\
\l^{i'}\frac{\partial}{\partial x^{i'}}&=&q^{-1}v\left (\beta\,\frac{\partial}{\partial\zeta}+\bar\beta\,\frac{\partial}{\partial\bar\zeta}\right )-q^{-1}F\,\frac{\partial}{\partial v}+q^{-1}\frac{\partial}{\partial u}\ , \nonumber \\
\label{16} \end{eqnarray}
we find that
\begin{equation}\label{17}
k_{i';j'}=-q^{-1}\bar\beta\,m_{i'}k_{j'}-q^{-1}\beta\,\bar m_{i'}k_{j'}\ ,
\end{equation}
where the semicolon denotes covariant differentiation with respect to the Riemannian connection calculated with the metric tensor given by the line element (\ref{12}). In general $\beta\neq 0$ and it follows from (\ref{2}) that the null hyperplanes $u={\rm constant}$ \emph{intersect} (having differing normals corresponding to different values of $u$). The waves in this case are Kundt waves and our geometrical construction, which is responsible for the introduction of the functions $\beta$ and $q$, demonstrates explicitly that the non-vanishing of the right hand side of (18) corresponds to intersecting wave fronts. On the other hand if $\beta=0$ (and thus $l={\rm constant}$ in (\ref{3})) the null hyperplanes $u={\rm constant}$ \emph{do not intersect} and the right hand side of (\ref{17}) vanishes so that $k_{i'}$ is covariantly constant. The waves in this case are plane fronted with parallel rays or pp--waves. In this case a simple transformation of the coordinate $u$ effectively makes $q=1$ in (\ref{12}) and (\ref{14}).

\section{Gravity Waves with Expanding Wave Fronts}\label{sec:gravwaves_expanding_wavefronts}

\begin{figure}
\begin{center}
\includegraphics[width=0.5\textwidth, trim= 0.1 5 0.1 10, clip=true]{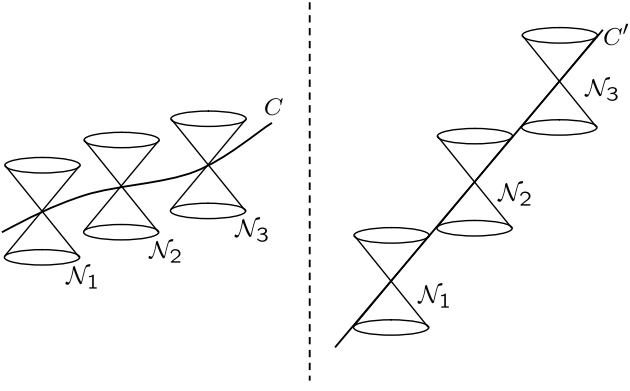}
\end{center}
  \caption{\label{fig_cones} In Minkowskian space-time we have on the lhs intersecting null cones ${\cal N}_1, {\cal N}_2, {\cal N}_3$ with vertices on an arbitrary world line $C$. On the rhs we have non-intersecting null cones ${\cal N}_1, {\cal N}_2, {\cal N}_3$ with vertices on a common generator (null geodesic) $C'$.}
\end{figure}

Shear--free null hypersurfaces in Minkowskian spacetime are, in addition to the null hyperplanes of the previous section, given by $u(X, Y, Z, T)={\rm constant}$ with the function $u(X, Y, Z, T)$ given implicitly by
\begin{equation}\label{18}
\eta_{ij}(X^i-w^i(u))(X^j-w^j(u))=0\ .
\end{equation}
In this case $u(X, Y, Z, T)={\rm constant}$, are null cones with vertices on the world line $X^i=w^i(u)$. This world line can be timelike, spacelike or null and $u$ is an arbitrary parameter along it. Differentiating (\ref{18}) partially with respect to $X^k$ results in
\begin{equation}\label{19}
u_{,k}=\frac{\xi_k}{R}\ \ \ {\rm with}\ \ \xi_k=\eta_{kl}(X^l-w^l(u))\ \ {\rm and}\ \ R=\eta_{ij}\dot w^i\,\xi^j\ .
\end{equation}
Clearly $u(X, Y, Z, T)={\rm constant}$ are null hypersurfaces and furthermore it is confirmed in \cite{Hogan:2018} that they are shear--free and expanding (or contracting). It is useful to write the four real--valued functions $w^i(u)$ in terms of a complex--valued function $l(u)$ (and its complex conjugate $\bar l(u)$), and two real--valued functions $m(u)$ and $n(u)$ according to 
\begin{equation}\label{20}
w^1+i\,w^2=-\frac{2\sqrt{2}\,l}{m}\ ,\ w^3+w^4=n-\frac{4\,l\,\bar l}{m}\ ,\ w^3-w^4=\frac{2}{m}\ .
\end{equation}
Comparing this with (\ref{3}) we note that
\begin{equation}\label{21}
\lim_{m\rightarrow 0}m\,\xi^i=a^i\ .
\end{equation}
The character of the world line $X^i=w^i(u)$ is important when considering, for future reference, whether or not the null hypersurfaces $u={\rm constant}$ (the null cones) intersect, c.f.\ Fig.\ \ref{fig_cones}. We see from (\ref{20}) that
\begin{equation}\label{1a}
\eta_{ij}\,\dot w^i\,\dot w^j=\frac{4\,\kappa}{m^2}\ \ \ {\rm with}\ \ \ \kappa=2\,|\beta|^2-\frac{1}{2}\alpha\,\gamma\ ,
\end{equation}
where $\beta=\dot l\ ,\ \alpha=\dot m$ and $\gamma=\dot n$. Thus the sign of $\kappa$ determines whether the world line $X^i=w^i(u)$ is timelike, spacelike or null. Clearly the hypersurfaces $u={\rm constant}$ intersect in general but they do not intersect if $X^i=w^i(u)$ is a common generator of the null cones. In this case $X^i=w^i(u)$ is a null geodesic and so 
\begin{equation}\label{2a}
\kappa=0\ \ \ {\rm and}\ \ \ \ddot w^i=C(u)\,\dot w^i\ ,
\end{equation}
for some real valued function $C$. With (\ref{20}) substituted into 
\begin{equation}\label{3a}
\ddot w^3-\ddot w^4=C(\dot w^3-\dot w^4)\ ,
\end{equation}
we find that
\begin{equation}\label{4a}
C=\frac{\dot\alpha}{\alpha}-\frac{2\,\alpha}{m}\ .
\end{equation}
Now
\begin{equation}\label{5a}
\ddot w^1+i\ddot w^2=C(\dot w^1+i\dot w^2)\ \ \ {\rm and}\ \ \ \kappa=0\ ,
\end{equation}
gives
\begin{equation}\label{6a}
\frac{\dot\beta}{\beta}=\frac{\dot\alpha}{\alpha}=\frac{\dot\gamma}{\gamma}\ ,
\end{equation}
and we note from the first equality here that $\dot\beta/\beta$ is real. Now the remaining equation
\begin{equation}\label{7a}
\ddot w^3+\ddot w^4=C(\dot w^3+\dot w^4)\ ,
\end{equation}
is automatically satisfied. We will return to these non--intersecting null hypersurface conditions (\ref{2a}) with (\ref{4a}) and (\ref{6a}) below.

In light of (\ref{20}) we can use (\ref{18}) to obtain
\begin{eqnarray}
Z+T&=&\sqrt{2}\,l\,(X-i\,Y)+\sqrt{2}\,\bar l\,(X+i\,Y)+2\,l\,\bar l\,(Z-T)\nonumber\\
&&+n\left (1-\frac{m}{2}(Z-T)\right )+\frac{m}{2}\,\eta_{ij}\,X^i\,X^j\ ,\label{22}
\end{eqnarray}
and thus we have
\begin{eqnarray}
&&\left (1-\frac{m}{2}(Z-T)\right )\eta_{ij}\,X^i\,X^j=\nonumber\\ 
&&|X+i\,Y+\sqrt{2}\,l(Z-T)|^2+n(Z-T)\left (1-\frac{m}{2}(Z-T)\right ). \nonumber \\ \label{23}
\end{eqnarray}
Consequently (\ref{22}) can take the form
\begin{eqnarray}
Z+T&=&\frac{m}{2}\left (1-\frac{m}{2}(Z-T)\right )^{-1} \nonumber\\
&&\times|X+i\,Y+\sqrt{2}\,l\,(Z-T)|^2+\sqrt{2}\,l\,(X-i\,Y)  \nonumber\\
&&+\sqrt{2}\,\bar l\,(X+i\,Y)+2\,l\,\bar l(Z-T)+n\,.  \label{24}
\end{eqnarray}
Now define
\begin{equation}\label{25}
\zeta =\left (1-\frac{m}{2}(Z-T)\right )^{-1}\left\{\frac{1}{\sqrt{2}}(X+i\,Y)+l(Z-T)\right\}\ ,
\end{equation}
and so we finally have
\begin{eqnarray}
X+i\,Y&=&\sqrt{2}\left (1-\frac{m}{2}(Z-T)\right )\zeta-\sqrt{2}\,l(Z-T)\ ,\label{26}\\
Z+T&=&2\left (1-\frac{m}{2}(Z-T)\right )\left (\bar l\,\zeta+l\,\bar\zeta+\frac{m}{2}\zeta\bar\zeta+\frac{n}{2}\right )\nonumber\\
&&-\left (2\,l\,\bar l-\frac{n\,m}{2}\right )(Z-T)\ .\label{27}
\end{eqnarray}
These equations reduce to (\ref{7}) and (\ref{8}) respectively when $m=0$. Substituting (\ref{26}) and (\ref{27}) into the Minkowskian line element (\ref{1}) we have
\begin{eqnarray}
ds_0^2&=&2\left (1-\frac{m}{2}(Z-T)\right )^2\left |d\zeta-\frac{(Z-T)\,q_{\bar\zeta}}{1-\frac{m}{2}(Z-T)}\,du\right |^2 \nonumber \\
&&+2\,q\,du\,(dZ-dT)\ ,\label{28}
\end{eqnarray}
with
\begin{equation}\label{29}
q(\zeta, \bar\zeta, u)=\bar\beta\,\zeta+\beta\,\bar\zeta+\frac{1}{2}\alpha\,\zeta\bar\zeta+\frac{1}{2}\gamma\ ,\ \ \ q_{\bar\zeta}=\frac{\partial q}{\partial\bar\zeta}\ .
\end{equation}

If we now substitute (\ref{20}), (\ref{26}) and (\ref{27}) into $R$ given in (\ref{19}) we find that
\begin{equation}\label{30}
R=\eta_{ij}\,\dot w^i\,\xi^j=-\frac{2}{m}\left (1-\frac{m}{2}(Z-T)\right )q\ ,
\end{equation}
with $q$ given by (\ref{29}). Thus
\begin{equation}\label{31}
\lim_{m\rightarrow 0}m\,R=-2\,q\ .
\end{equation}
Writing the first of (\ref{19}) as the equality of 1--forms
\begin{equation}\label{32}
m\,\xi_i\,dX^i=m\,R\,du\ ,
\end{equation}
we see from (\ref{21}) and (\ref{31}) that this equation becomes 
\begin{equation}\label{33}
a_i\,dX^i=-2\,q\,du\ ,
\end{equation}
in the limit $m\rightarrow 0$, thereby recovering (\ref{11}) as a special case. To model expanding gravitational waves propagating in a vacuum with a solution of Einstein's vacuum field equations we require a metric tensor of the form $g_{ij}=\eta_{ij}+\lambda_i\,\xi_j+\lambda_j\,\xi_i$ for some covariant vector $\lambda_i$. This will result in $g_{ij}\,\xi^i\,\xi^j=\eta_{ij}\,\xi^i\,\xi^j=0$ so the equation for $u(X, Y, Z, T)$ will be identical in Minkowskian spacetime and in the space--time with metric $g_{ij}$. This form of $g_{ij}$ can be achieved by working with the Robinson--Trautman \cite{Robinson:Trautman:1960,Robinson:Trautman:1962} form of metric. From now on it is convenient to work in the coordinates $x^{i'}=(\zeta, \bar\zeta, Z-T, u)$ instead of the rectangular Cartesians and time $X^i$. Thus our first task is to choose $Z-T$ in (\ref{28}) so that the Minkowskian line element (\ref{28}) assumes Robinson--Trautman form. This is achieved simply by introducing a coordinate $r$ via the equation
\begin{equation}\label{34}
Z-T=\frac{2}{m}-\frac{r}{q}\ .
\end{equation}
This has the desired effect of turning (\ref{28}) into 
\begin{equation}\label{35}
ds_0^2=2\,r^2p_0^{-2}|d\zeta+Q_0\,du|^2-2\,du\,dr-(K_0-2\,r\,H_0)du^2,
\end{equation}
with
\begin{eqnarray}
p_0&=&\frac{2\,q}{m}\ ,\ Q_0=\frac{2\,q_{\bar\zeta}}{m}\,\left (\Rightarrow \frac{\partial Q_0}{\partial\bar\zeta}=0\right )\ , \nonumber \\
 K_0&=&-\frac{4\,\kappa}{m^2}=\Delta\log q\ ,\label{36}
\end{eqnarray}
and
\begin{eqnarray}
H_0&=&q^{-1}\dot q-\frac{4}{m}\,q^{-1}q_{\zeta}\,q_{\bar\zeta}\ ,\nonumber\\
&=&p_0^{-1}\dot p_0+\frac{1}{2}p_0^2\frac{\partial}{\partial\zeta}(p_0^{-2}Q_0)+\frac{1}{2}p_0^2\frac{\partial}{\partial\bar\zeta}(p_0^{-2}\bar Q_0).\label{37}
\end{eqnarray}
Here
\begin{equation}\label{38}
\Delta=2\,p_0^2\frac{\partial^2}{\partial\zeta\partial\bar\zeta}\ .
\end{equation}
Now following Robinson and Trautman the spacetime model of expanding gravitational radiation propagating in a vacuum is given by the line element
\begin{equation}\label{39}
ds^2=2\,r^2p^{-2}|d\zeta+Q\,du|^2-2\,du\,dr-(K-2\,r\,H)du^2\ ,
\end{equation}
with
\begin{equation}\label{40}
p=p_0\ ,\ Q=Q_0+G(\zeta, u)\ ,\ K=K_0\ ,
\end{equation}
and
\begin{equation}\label{41}
H=H_0+\frac{1}{2}p_0^2\frac{\partial}{\partial\zeta}(p_0^{-2}G)+\frac{1}{2}p_0^2\frac{\partial}{\partial\bar\zeta}(p_0^{-2}\bar G)\ ,
\end{equation}
where $G(\zeta, u)$ is an arbitrary analytic function. The Newman--Penrose \cite{Newman:Penrose:1962} components $\Psi_A$ with $A=0, 1, 2, 3, 4$, of the Riemann curvature tensor vanish with the exception of 
\begin{equation}\label{42}
\Psi_4=\frac{1}{r}\frac{\partial}{\partial\zeta}\left (p_0^2\frac{\partial}{\partial\zeta}(H-H_0)\right )\ ,
\end{equation}
indicating Petrov Type N with degenerate principal null direction $\partial/\partial r$. 

The null hypersurfaces $u={\rm constant}$, in the spacetime with line element (\ref{39}), \emph{intersect} in general. To impose the conditions (\ref{2a}) with (\ref{4a}) and (\ref{6a}) we first note that $H_0$ in (\ref{37}) can be written out explicitly in the form
\begin{eqnarray}
H_0&=&q^{-1}\Biggl\{\left (\frac{\dot{\bar\beta}}{\bar\beta}-\frac{\dot\alpha}{\alpha}\right )\bar\beta\,\zeta+\left (\frac{\dot{\beta}}{\beta}-\frac{\dot\alpha}{\alpha}\right )\beta\bar\zeta \nonumber \\
&&+\frac{1}{2}\left (\frac{\dot\gamma}{\gamma}-\frac{\dot\alpha}{\alpha}\right )\gamma-\frac{2\,\kappa}{m}+C\,q\Biggr\}\ ,\label{43}
\end{eqnarray}
with $\kappa$ given by (\ref{1a}) and $C$ given by (\ref{4a}). If we require that the null hypersurfaces $u={\rm constant}$ are \emph{non--intersecting} then $H_0=C$ and, from (\ref{36}), $K_0=0$. In addition if we choose $u$ to be an affine parameter along the null geodesic (\ref{2a}) then $C=0$ and thus $H_0=0$. In the non--intersecting case with $C=0$ the coordinate transformation
\begin{equation}\label{44}
\zeta'=-\frac{m}{2}\left (\beta+\frac{1}{2}\alpha\,\zeta\right )^{-1}\ ,
\end{equation}
leads to
\begin{equation}\label{44'}
d\zeta=\frac{du}{\zeta'}+\frac{m}{\alpha\,\zeta'^2}\,d\zeta'\ ,\ p_0=\frac{m}{\alpha\,\zeta'\bar\zeta'}\ \ {\rm and}\ \ Q_0=-\frac{1}{\zeta'}\ ,
\end{equation}
and as a result
\begin{equation}\label{45'}
d\zeta+(Q_0+G)\,du=\frac{m}{\alpha\,\zeta'^2}(d\zeta'+G'du)\ ,
\end{equation}
with
\begin{equation}\label{46'}
G'(\zeta', u)=\frac{\alpha\,\zeta'^2}{m}\,G(\zeta, u)\ ,
\end{equation}
and $\zeta$ in the argument of the function $G$ replaced by $\zeta'$ using (\ref{44}). Now
\begin{equation}\label{45}
p_0^{-2}|d\zeta+(Q_0+G)\,du|^2=|d\zeta'+G'du|^2\ .
\end{equation}
Also $H$ in (\ref{41}) with $C=0$ reads
\begin{equation}\label{47'}
H=\frac{1}{2}\left (\frac{\partial G'}{\partial\zeta'}+\frac{\partial\bar G'}{\partial\bar\zeta'}\right )\ ,
\end{equation}
since
\begin{equation}\label{48'}
p_0^{-2}G=\frac{\alpha\,\bar\zeta'^2}{m}\,G'\ \ \ {\rm and}\ \ \ p_0^2\,\frac{\partial}{\partial\zeta}=\frac{m}{\alpha\,\bar\zeta'^2}\,\frac{\partial}{\partial\zeta'}\ .
\end{equation}
Consequently in the non--intersecting case the line element (\ref{39}) can be written in the form \cite{Robinson:Trautman:1960,Robinson:Trautman:1962}
\begin{eqnarray}
ds^2&=&2\,r^2|d\zeta'+G'(\zeta', u)\,du|^2-2\,du\,dr \nonumber \\
&&+r\,\left (\frac{\partial G'}{\partial\zeta'}+\frac{\partial\bar G'}{\partial\bar\zeta'}\right )du^2\ ,\label{46}
\end{eqnarray}
and
\begin{equation}\label{47}
\Psi_4=\frac{1}{2\,r}\frac{\partial^3G'}{\partial\zeta'^3}\ .
\end{equation}

\section{Discussion}\label{sec:discussion}

The radiative solutions of Einstein's vacuum field equations utilized in this paper play an important role in understanding aspects of gravitational waves in general. They have recently been shown to have gravitational fields (i.e.\ Riemann tensors) which, in a precise technical sense, are proportional to the square of a radiative Maxwell field  \cite{Godazgar:etal:2021}. The Kundt waves and the pp--waves are usually described by two distinct line elements. However a byproduct of section \ref{sec:planewaves} is a line element (\ref{12}) which incorporates both cases depending upon the choice of the arbitrary complex--valued function $\beta(u)$. This line element can be written in a form which is closer in form to that of Kundt by making the coordinate transformation $v\rightarrow\ v'=q\,v$ resulting in
\begin{equation}\label{48}
ds^2=2\,d\zeta\,d\bar\zeta+2\,du\,(dv'+\bar W\,d\zeta+W\,d\bar\zeta+G\,du)\ ,
\end{equation}
with
\begin{equation}\label{49}
W=-2\,q^{-1}\beta\,v'\ \ {\rm and}\ \  G=q^{-2}\beta\,\bar\beta\,v'^2-q^{-1}\,\dot q\,v'+q\,F\ ,
\end{equation}
with $\dot q=\partial q/\partial u$. Here the function $q$ is given by (\ref{10}). The explicit terms in (\ref{49}) involving the functions $\beta$ and $q$ are unique to our geometrical construction given in section \ref{sec:planewaves}. The waves given by Kundt correspond to the choice $q=\zeta+\bar\zeta$ so that $\beta=1$ and $\gamma=0$ in (\ref{10}).

In conclusion, we provided a scheme which allows for an easy identification whether or not the wave fronts of gravitational radiation intersect. Furthermore, we found a line element which allows for the unified description of Kundt and pp-waves.

\begin{acknowledgments}
This work was funded by the Deutsche Forschungsgemeinschaft (DFG, German Research Foundation) through the grant PU 461/1-2 -- project number 369402949 (D.P.). 
\end{acknowledgments}

\appendix

\section{Plane Fronted Electromagnetic Waves}\label{app_A}

In this appendix we demonstrate how our geometrical construction of plane fronted gravitational waves in section \ref{sec:planewaves} can be utilized to describe plane fronted electromagnetic waves in the context of Minkowskian space--time \cite{Hogan:Puetzfeld:2021}. We show how these waves are (a) members of the family of Bateman electromagnetic waves, (b) identify the wave velocity and the angular velocity with which the searchlight beam described by Pirani sweeps across the sky, (c) confirm the radiative character of the electromagnetic field and (d) outline the polarization properties of the waves

The line element of Minkowskian space--time reads (\ref{9})
\begin{equation}\label{A1}
ds^2=\eta_{ij}\,dX^i\,dX^j=2\,|d\zeta-\beta(u)\,v\,du|^2+2\,q\,du\,dv\ ,
\end{equation}
with $X^i=(X, Y, Z, T)$ and $\eta_{ij}={\rm diag}(1, 1, 1, -1)$. Plane fronted electromagnetic waves propagating in a vacuum are described by a potential 1--form
\begin{equation}\label{A2}
A=f(\zeta, \bar\zeta, u)\,du\ ,
\end{equation}
where $f(\zeta, \bar\zeta, u)$ is a real--valued function of its arguments. The corresponding Maxwell 2--form is given by the exterior derivative
\begin{equation}\label{A3}
F=dA=\frac{\partial f}{\partial\zeta}\,d\zeta\wedge du+\frac{\partial f}{\partial\bar\zeta}\,d\bar\zeta\wedge du\ ,
\end{equation}
and Maxwell's vacuum field equations require
\begin{equation}\label{A4}
\frac{\partial^2f}{\partial\zeta\partial\bar\zeta}=0\ .
\end{equation}
Hence $\partial f/\partial\zeta=g(\zeta, u)$ is an arbitrary complex--valued analytic function. Then the Maxwell 2--form gives the complex--valued 2--form
\begin{equation}\label{A5}
{\cal F}:=F-i\,{}^*F=2\,g(\zeta, u)\,d\zeta\wedge du\ ,
\end{equation}
where the star denotes the Hodge dual and we have
\begin{equation}\label{A6}
{}^*{\cal F}=i\,{\cal F}\ .
\end{equation}
We will express this Maxwell field in terms of the coordinates $X^i=(X, Y, Z, T)$ and then read off the corresponding electric 3--vector ${\bf E}$ and magnetic 3--vector ${\bf B}$. In addition we will show that these electromagnetic waves are examples of Bateman's wave solutions of Maxwell's vacuum field equations.

Using equations (\ref{A5}), (\ref{A6}) and (\ref{11}) we have
\begin{eqnarray}
\zeta&=&\frac{1}{\sqrt{2}}(X+i\,Y)+l(u)\,(Z-T)\ , \label{A7} \\
d\zeta&=&\zeta_{,i}\,dX^i=\frac{1}{\sqrt{2}}(dX+i\,dY)+l(u)\,(dZ-dT)\nonumber \\
&&+\beta(u)\,(Z-T)\,du\ , \label{A8}
\end{eqnarray}
and
\begin{eqnarray}
du&=&-\frac{1}{2\,q}\,a_i\,dX^i\ ,\nonumber\\
&=&-\frac{1}{2\,q}\Biggl\{\sqrt{2}\,l(u)\,(dX-i\,dY)+\sqrt{2}\,\bar l(u)\,(dX+i\,dY)\nonumber\\
&&+2\,l(u)\,\bar l(u)\,(dZ-dT)-(dZ+dT)\Biggr\}\ .\label{A9}
\end{eqnarray}
Hence we can write (\ref{A5}) as
\begin{eqnarray}
{\cal F}&=&q^{-1}g(\zeta, u)\Biggl\{\frac{(1+2\,l^2)}{\sqrt{2}}(dX\wedge dZ+i\,dY\wedge dT)\nonumber\\
&&+\frac{i\,(1-2\,l^2)}{\sqrt{2}}(dY\wedge dZ-i\,dX\wedge dT)\nonumber \\
&&+2\,i\,l\,(dX\wedge dY-i\,dZ\wedge dT)\Biggr\}\ .\nonumber\\\label{A10}
\end{eqnarray}
We note that using the dual 2--forms
\begin{eqnarray}
{}^*(dX\wedge dY)&=&dZ\wedge dT,\ {}^*(dZ\wedge dX)=dY\wedge dT,\nonumber \\ 
{}^*(dY\wedge dZ)&=&dX\wedge dT,\label{A11}
\end{eqnarray}
and the fact that the dual of the dual results in the original 2--form multiplied by minus one (so that, for example, ${}^*(dZ\wedge dT)=-dX\wedge dY$) we easily confirm that (\ref{A10}) satisfies (\ref{A6}). It is understood that $\zeta$ in the argument of the function $g$ in (\ref{A10}) is expressed in terms of $X, Y, Z, T$ via (\ref{7}). To write (\ref{A10}) in the form of Bateman's waves we first need the following equations satisfied by $\zeta(X, Y, Z, T)$ and $u(X, Y, Z, T)$ involving Jacobian determinants:
\begin{eqnarray}
\frac{\partial(\zeta, u)}{\partial(Z, X)}&=&i\,\frac{\partial(\zeta, u)}{\partial(Y, T)}=-\frac{(1+2\,l^2)}{2\,\sqrt{2}\,q}\ ,\label{A12}\\
\frac{\partial(\zeta, u)}{\partial(Y, Z)}&=&i\,\frac{\partial(\zeta, u)}{\partial(X, T)}=\frac{i\,(1-2\,l^2)}{2\,\sqrt{2}\,q}\ ,\label{A13}\\
\frac{\partial(\zeta, u)}{\partial(X, Y)}&=&i\,\frac{\partial(\zeta, u)}{\partial(Z, T)}=\frac{i\,l}{q}\ .\label{A14}
\end{eqnarray}
Consequently we can write (\ref{A10}) in the form
\begin{eqnarray}
{\cal F}&=&-2\,g\,\frac{\partial(\zeta, u)}{\partial(X, T)}\,dT\wedge dX \nonumber \\
&&-2\,g\,\frac{\partial(\zeta, u)}{\partial(Y, T)}\,dT\wedge dY-2\,g\,\frac{\partial(\zeta, u)}{\partial(Z, T)}\,dT\wedge dZ\nonumber\\
&&+2\,g\,\frac{\partial(\zeta, u)}{\partial(Y, Z)}\,dY\wedge dZ+2\,g\,\frac{\partial(\zeta, u)}{\partial(X, Y)}\,dX\wedge dY\nonumber \\
&&+2\,g\,\frac{\partial(\zeta, u)}{\partial(Z, X)}\,dZ\wedge dX\ .\nonumber\\\label{A15}
\end{eqnarray}
We note that this complex--valued 2--form is thus given by
\begin{equation}\label{A16}
{\cal F}=\frac{1}{2}{\cal F}_{ij}\,dX^i\wedge dX^j\ \ {\rm with}\ \ {\cal F}_{ij}=2\,g(\zeta, u)\,(\zeta_{,i}\,u_{,j}-\zeta_{,j}\,u_{,i})\ ,
\end{equation}
confirming (\ref{A5}). With $\zeta_{,i}$ and $u_{,i}$ given by (\ref{A8}) and (\ref{A9}) respectively we see now that
\begin{equation}\label{A17}
{\cal F}_{ij}\,a^j=0\ ,
\end{equation}
confirming that $a^i$ is a degenerate principle null direction of this Maxwell field. If in coordinates $X^i$ the electric 3--vector has components ${\bf E}=(E^1, E^2, E^3)$ and the magnetic 3--vector has components ${\bf B}=(B^1, B^2, B^3)$ then these components can be read off from ${\cal F}$ using
\begin{eqnarray}
{\cal F}&=&-i\,(B^1+i\,E^1)\,dT\wedge dX-i\,(B^2+i\,E^2)\,dT\wedge dY\nonumber\\
&&-i\,(B^3+i\,E^3)\,dT\wedge dZ-(B^1+i\,E^1)\,dY\wedge dZ\nonumber\\
&&-(B^3+i\,E^3)\,dX\wedge dY-(B^2+i\,E^2)\,dZ\wedge dX\ .\nonumber \\\label{A18}
\end{eqnarray}
From (\ref{A15}) and (\ref{A18}) we arrive at the Bateman \cite[p.12 (written in 1912)]{Bateman:1955} form of this Maxwell field namely, 
\begin{eqnarray}
B^1+i\,E^1&=&-2\,g\,\frac{\partial(\zeta, u)}{\partial(Y, Z)}\ ,\label{A18a}\\
B^2+i\,E^2&=&-2\,g\,\frac{\partial(\zeta, u)}{\partial(Z, X)}\ ,\label{A18b}\\
B^3+i\,E^3&=&-2\,g\,\frac{\partial(\zeta, u)}{\partial(X, Y)}\ ,\label{A18c}\end{eqnarray} 
together with the first equations in (\ref{A12})--(\ref{A14}). Since we get two expressions for each of the components of ${\bf B}+i\,{\bf E}$ when comparing (\ref{A15}) with (\ref{A18}), the consistency of these expressions follows from the equations (\ref{A12})--(\ref{A14}). Writing
\begin{equation}\label{A19}
\varphi=\frac{1}{\sqrt{2}\,q}\,g(\zeta, u)\ ,
\end{equation}
we have
\begin{eqnarray}
B^1+i\,E^1&=&-i\,\varphi\,(1-2\,l^2)\ ,\ B^2+i\,E^2=\varphi\,(1+2\,l^2)\ , \nonumber \\
B^3+i\,E^3&=&-2\,\sqrt{2}\,\varphi\,i\,l\ .\label{A20}
\end{eqnarray}
These can be written neatly in 3--vector form as
\begin{equation}\label{x1}
{\bf B}+i\,{\bf E}=-i\,\varphi\,\sqrt{2}\,(2\,l\,\bar l+1){\bf m}\ ,
\end{equation}
with 
\begin{equation}\label{x2}
{\bf m}=\left (\frac{1-2\,l^2}{\sqrt{2}\,(2\,l\,\bar l+1)}, \frac{i(1+2\,l^2)}{\sqrt{2}\,(2\,l\,\bar l+1)}, \frac{2\,l}{(2\,l\,\bar l+1)}\right )\ .
\end{equation}

The histories of the plane wave fronts are $u(X, Y, Z, T)={\rm constant}$. Consequently the \emph{wave velocity} (see \cite[p.418]{Synge:1965}) has components
\begin{equation}\label{A22}
v^{\alpha}=-\frac{u_{,4}\,u_{,\alpha}}{u_{,\beta}\,u_{,\beta}}\ ,
\end{equation}
with Greek indices taking values 1, 2, 3. Since $u_{,i}=-(2\,q)^{-1}a_i$ we can write the wave velocity as
\begin{equation}\label{A23}
{\bf v}=(a^4)^{-1}{\bf a}\ ,
\end{equation}
and this is a unit 3--vector (${\bf v}\cdot{\bf v}=1$) confirming that the waves travel with the speed of light. In general $d{\bf v}/du\neq 0$ for intersecting plane fronted waves while for non--intersecting plane fronted waves ${\bf v}$ is a constant 3--vector ($\Leftrightarrow\ \beta(u)=0$). With ${\bf v}$ given by (\ref{23}) and ${\bf m}$ (with complex conjugate denoted $\bar{\bf m}$) by (\ref{x2}) we find that
\begin{equation}\label{A25}
{\bf m}\cdot \bar{\bf m}=1\ ,\ {\bf m}\cdot{\bf m}=0\ ,\ \bar{\bf m}\cdot\bar{\bf m}=0\ ,\ {\bf m}\times\bar{\bf m}=i\,{\bf v}\end{equation}
and thus
\begin{equation}\label{A26}
{\bf m}\times{\bf v}=-i\,{\bf m}\ ,\bar{\bf m}\times {\bf v}=i\,\bar{\bf m}\ ,\ {\bf m}\cdot{\bf v}=0\ ,\ \bar{\bf m}\cdot{\bf v}=0\ .\end{equation}
We see from (\ref{A9}) that
\begin{equation}\label{x3}
\frac{\partial u}{\partial T}=\frac{(2\,l\,\bar l+1)}{2\,q}\ ,\end{equation}
and so we find that
\begin{equation}\label{A27}
\frac{d{\bf v}}{dT}=\frac{(2\,l\,\bar l+1)}{2\,q}\frac{d{\bf v}}{du}=q^{-1}\,\bar\beta\,{\bf m}+q^{-1}\,\beta\,\bar{\bf m}\ .\end{equation}
Using the first two equations in (\ref{A26}) we can write this as
\begin{equation}\label{x4}
\frac{d{\bf v}}{dT}={\boldsymbol\omega}\times{\bf v}\ ,\end{equation}
with
\begin{equation}\label{A28}
{\boldsymbol\omega}=i\,q^{-1}\bar\beta\,{\bf m}-i\,q^{-1}\beta\,\bar{\bf m}\ \ \ \left (\Rightarrow\ \ {\boldsymbol\omega}\cdot{\boldsymbol\omega}=2\,q^{-2}|\beta|^2\right )\ ,
\end{equation}
and thus \emph{${\boldsymbol\omega}$ is the angular velocity with which the searchlight beam described by Pirani sweeps across the sky}, with ${\boldsymbol\omega}=0\ \Leftrightarrow\ \beta=0$.

Starting with ${\bf B}+i\,{\bf E}$ given by (\ref{x1}) we see that ${\bf m}\cdot{\bf m}=0$ implies
\begin{equation}\label{x5}
{\bf E}\cdot{\bf E}={\bf B}\cdot{\bf B}\ \ {\rm and}\ \ {\bf E}\cdot{\bf B}=0\ ,
\end{equation}
\emph{confirming the radiative character of the electromagnetic field}. Since ${\bf m}\cdot{\bf v}=0$ we have
\begin{equation}\label{A29}
{\bf E}\cdot{\bf v}=0={\bf B}\cdot{\bf v}\ .
\end{equation}
With ${\bf m}\cdot\bar{\bf m}=1$, and $|{\bf E}|=|{\bf B}|$ on account of (\ref{x5}),
\begin{equation}\label{x6}
|{\bf E}|=|{\bf B}|=|\varphi |(2\,l\,\bar l+1)\ ,
\end{equation}
and so writing $\varphi=|\varphi |\,e^{i\,\vartheta}$ we have (\ref{x1}) in the form
\begin{equation}\label{x7}
\frac{1}{\sqrt{2}}({\bf E}-i\,{\bf B})=-|{\bf E}|\,e^{i\,\vartheta}{\bf m}\ .
\end{equation}
Now ${\bf m}\times{\bf v}=-i\,{\bf m}$ applied to this yields
\begin{equation}\label{x8}
\frac{1}{\sqrt{2}}({\bf E}\times{\bf v}-i\,{\bf B}\times{\bf v})=i\,|{\bf E}|\,e^{i\,\vartheta}{\bf m}=\frac{1}{\sqrt{2}}(-{\bf B}-i\,{\bf E})\ ,
\end{equation}
from which we conclude that
\begin{equation}\label{x9}
{\bf v}\times{\bf E}={\bf B}\ \ \ {\rm and}\ \ \ {\bf B}\times{\bf v}={\bf E}\ .
\end{equation}
Finally using ${\bf m}\times\bar{\bf m}=i\,{\bf v}$ we obtain from (\ref{x7}) that
\begin{equation}\label{x10}
\frac{1}{2}({\bf E}-i\,{\bf B})\times({\bf E}+i\,{\bf B})=|{\bf E}|^2{\bf m}\times\bar{\bf m}=i\,|{\bf E}|^2{\bf v}\ ,
\end{equation}
from which we arrive at the Poynting vector
\begin{equation}\label{x11}
{\bf E}\times{\bf B}=|{\bf E}|^2{\bf v}\ .
\end{equation}
From the second of (\ref{x5}) along with (\ref{A29}), (\ref{x9}) and (\ref{x11}) we see that \emph{${\bf E}, {\bf B}, {\bf v}$ constitute a right handed triad}. Also since (\ref{x11}) demonstrates that the Poynting vector points in the direction of the wave velocity ${\bf v}$ it follows that \emph{the energy of the waves is propagated with velocity ${\bf v}$}.

To obtain information on \emph{the polarisation of the electromagnetic waves} we require the rate of change of ${\bf E}$ and of ${\bf B}$ with respect to the coordinate time $T$. We can obtain this from (\ref{x7}) with the help of
\begin{equation}\label{x12}
\frac{\partial {\bf m}}{\partial T}=-q^{-1}\beta\,{\bf v}+q^{-1}(\beta\,\bar l-\bar\beta\,l)\,{\bf m}\ .
\end{equation}
With this we find that
\begin{eqnarray}
\frac{\partial {\bf E}}{\partial T}&=&\frac{1}{|{\bf E}|}\frac{\partial |{\bf E}|}{\partial T}\,{\bf E}+\left (\frac{\partial\vartheta}{\partial T}-i\,q^{-1}(\beta\,\bar l-\bar\beta\,l)\right ){\bf v}\times{\bf E}\nonumber \\
&&-({\boldsymbol\omega}\cdot{\bf B}){\bf v}\ ,\label{x13}
\end{eqnarray}
and
\begin{eqnarray}
\frac{\partial {\bf B}}{\partial T}&=&\frac{1}{|{\bf B}|}\frac{\partial |{\bf B}|}{\partial T}\,{\bf B}+\left (\frac{\partial\vartheta}{\partial T}-i\,q^{-1}(\beta\,\bar l-\bar\beta\,l)\right ){\bf v}\times{\bf B}\nonumber\\
&&+({\boldsymbol\omega}\cdot{\bf E}){\bf v}\ .\label{x14}
\end{eqnarray}
We note that (\ref{x14}) is a consequence of the vector product of ${\bf v}$ with (\ref{x13}). Equations (\ref{x13}) and (\ref{x14}) can be written more simply as
\begin{equation}\label{x15}
\frac{\partial}{\partial T}\left (\frac{{\bf E}}{|{\bf E}|}\right )=\left\{\frac{\partial\vartheta}{\partial T}-i\,q^{-1}(\beta\,\bar l-\bar\beta\,l)\right\}{\bf v}\times\frac{{\bf E}}{|{\bf E}|}+\boldsymbol\omega\times\frac{{\bf E}}{|{\bf E}|}\ ,
\end{equation}
and
\begin{equation}\label{x16}
\frac{\partial}{\partial T}\left (\frac{{\bf B}}{|{\bf B}|}\right )=\left\{\frac{\partial\vartheta}{\partial T}-i\,q^{-1}(\beta\,\bar l-\bar\beta\,l)\right\}{\bf v}\times\frac{{\bf B}}{|{\bf B}|}+\boldsymbol\omega\times\frac{{\bf B}}{|{\bf B}|}\ 
\end{equation}
together with 
\begin{equation}\label{x17}
\frac{\partial |{\bf E}|}{\partial T}=\left\{\frac{\partial}{\partial T}\log|{\varphi}|+q^{-1}(l\,\bar\beta+\bar l\,\beta)\right\}|{\bf E}|\ ,
\end{equation}
where we have made use of (\ref{x6}) in deriving the last equation here. These equations are clearly generalisations of plane polarisation for which the right hand sides of (\ref{x15}) and (\ref{x16}) vanish (so that the electric and magnetic vectors are fixed in direction) and circular polarisation for which the right hand side of (\ref{x17}) vanishes while (\ref{x15}) and (\ref{x16}) take the reduced form
\begin{equation}\label{x18}
\frac{\partial {\bf E}}{\partial T}=C\,{\bf v}\times{{\bf E}}\ \ \ {\rm and}\ \ \ \frac{\partial{\bf B}}{\partial T}=C\,{\bf v}\times{\bf B}\ ,
\end{equation}
where $C$ is a real constant. The final term in (\ref{x15}) and in (\ref{x16}) represents the rotation, with angular velocity $\boldsymbol\omega$, of the electric and magnetic vectors necessary to preserve the orthogonality of the triad ${\bf E}, {\bf B}, {\bf v}$ on account of the rotation of the wave velocity described in (\ref{x4}).

\bibliographystyle{unsrtnat}
\bibliography{colliding_bibliography}
\end{document}